\documentclass[preprint,12pt]{elsarticle}

\usepackage{amssymb}
\usepackage{epstopdf}

\journal{Applied Mathematics and Computation}

\begin{document}

\begin{frontmatter}



\newtheorem{thm}{Theorem}
\newproof{pf}{Proof}

\title{Efficient Approaches to the Mixture Distance Problem}



\author[fn1]{Justie Su-Tzu Juan}
\address[fn1]{Department of Computer Science and Information Engineering, National Chi Nan University, Puli, Nantou 54561 Taiwan}

\author[fn2]{Yi-Ching Chen}
\address[fn2]{Department of Computer Science and Information Engineering, National Taiwan University, Taipei, 10617 Taiwan}

\author[fn1]{Chen-Hui Lin}

\author[fn3]{Shu-Chuan Chen\corref{cor1}}
\cortext[cor1]{Corresponding author}
\ead{scchen@isu.edu}
\address[fn3]{Department of Mathematics and Statistics, Idaho State University, Pocatello, ID 83209 USA}



\begin{abstract}
Ancestral mixture model, an important model building a hierarchical tree from high dimensional binary sequences, was proposed by Chen and
Lindsay in 2006. As a phylogenetic tree (or evolutionary tree), a mixture tree created from ancestral mixture models, involves in the inferred
evolutionary relationships among various biological species. Moreover, it contains the information of time
when the species mutates. Tree comparison metric, an essential issue in bioinformatics, is to measure the
similarity between trees. To our knowledge, however, the approach to the comparison between two mixture trees is
still unknown. In this paper, we propose a new metric, named mixture distance metric, to measure the
similarity of two mixture trees. It uniquely considers the factor of evolutionary times between trees.
In addition, we further develop two algorithms to compute the mixture distance between
two mixture trees. One requires $O(n^2)$ and the other requires $O(n h)$ computation time with $O(n)$ preprocessing time, where $n$ denotes the number of leaves in the two mixture trees, and $h$ denotes the minimum height of these two trees.
\end{abstract}


\end{frontmatter}


\section{Introduction}\label{intro}
\begin{figure}[h]\centering
    \includegraphics[scale=0.8]{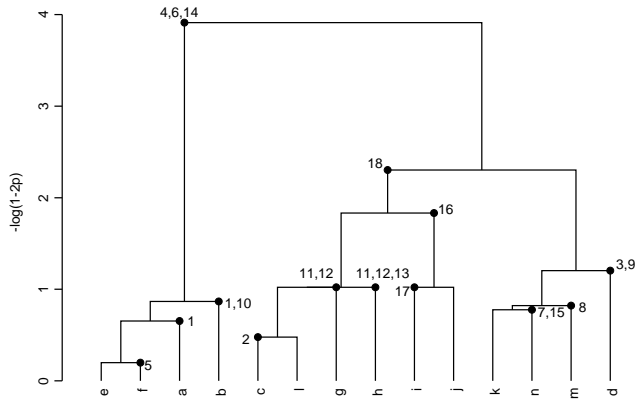}
    \caption{An example of the mixture tree~\cite{CL2006}.}
        \label{fig_mixture}
    \end{figure}

Phylogeny reconstruction involves reconstructing the evolutionary relationship from biological sequences among species. Nowadays it has become a critical issue in molecular biology and bioinformatics. Several
existing methods, such as neighbor-joining methods ~\cite{SN1987} and maximum likelihood methods ~\cite{LK1992} have been proposed to reconstruct a phylogenetic tree.  A novel and natural method, ancestral mixture models~\cite{CL2006}, was developed by Chen and Lindsay to deal with such a problem.
Mixture tree, a hierarchical tree created from ancestral mixture model, induces a sieve parameter to represent the evolutionary time.  Chen, Rosenberg and Lindsay (2011) then developed MixtureTree algorithm~\cite{CSL2011}, a linux based program written in C++, employed the ancestral mixture models to reconstruct mixture tree from DNA sequences.  With the information provided by the mixture tree,
 one can identify when and how a mutation event of species occurs. An example of the mixture tree created by MixtureTree algorithm ~\cite{CSL2011} is shown in Fig.~\ref{fig_mixture}. The data from Griffiths and Tavare (1994)~\cite{GT1994} are a subset of the mitochondrial DNA sequences which first appeared in Ward et al. (1991) ~\cite{Ward1991}.
 It is to study the mitochondrial diversity within the Nuu-Chuah-Nulth, an
Amerindian tribe from Vancouver Island. Ward et al. (1991) ~\cite{Ward1991} sequenced 360 nucleotide
segments of the mitochondrial control region for 63 individuals from the Nuu-Chuah-
Nulth. Griffiths and Tavare's subsample consisted of 55 of the 63 distinct sequences and 18
segregating sites including 13 pyrimidines (C, T ) and 5 purines (A, G). Each linage represents a distinct sequence, that is there are lineages $a$ through $n$. The time scale on the tree can be represented by $-\log(1-2p)$, where $p$ is a parameter, the mutation rate. The number on the tree represents the site of the lineage that the mutation occurs. For example, when $p=0.01$, lineages $e$ and $f$ merge because mutation occurs at site 5 of lineage $f$.

 Distinct methods may produce distinct trees, even though the methods adopt an identical
dataset~\cite{Steel1994}. To uncover a well-represented tree involving in evolutionary relationship among
species it is quite important to estimate how similar (or different) are among these trees. The tree
distance between two trees is a general measurement for the similarity of the trees.

Tree distance problem is a traditional issue in mathematics. Several metrics have been proposed to measure
the similarity between two trees, such as partition metric~\cite{RF1981}, quartet metric~\cite{EMM1985},
nearest neighbour interchange metric~\cite{DHJLTZ2000} and nodal distance metric~\cite{BS2003}. Because those metrics all compare two trees by considering the tree structure only, and does not mention about any parameter in the tree. So, those metrics are not suitable for  computing the similarity between two mixture trees. Therefore, we propose a novel metric, named mixture distance metric, to measure the similarity of two mixture trees in this paper. Among above previous metrics, the metric from the nodal distance algorithm is similar to our proposed metric. In 2003, John Bluis and Dong-Guk Shin~\cite{BS2003} presented the nodal distance algorithm which is used to measure the distance from leaves to other leaves for all leaves in a tree. The metric is defined as follows: Distance$(T_1, T_2) = \sum_{x, y \in L(T_1)=L(T_2)}|D_{T_1}(x, y) - D_{T_2}(x, y)|$. Where $D_{T_i}(x, y)$ denotes the distance of leaf $x$ to leaf $y$ in the tree $T_i$. The nodal distance algorithm was developed for this metric. Anyway, using this metric to measure the distance between two mixture trees is not conformable.

Over the metric of the mixture distance, the time parameter indicating when a mutation event of species occurs plays an important role in the tree similarity, which is however not considered by those previous metrics. We further develop two algorithms to compute the mixture distance between two mixture trees. One requires $O(n^2)$ and the other requires $O(n h)$ computation time with $O(n)$ preprocessing time, where $n$ denotes the number of leaves in these two mixture trees, and $h$ denoted the minimum height of these two trees. If we use the nodal distance algorithm with mixture distance metric, the time complexity will be $O(n^3)$ for binary unrooted trees. Comparisons with the methods perform on nodal distance show our methods perform better.

\section{Mixture Distance Metric}

A tree $T = (V(T), E(T))$ is a connected and acyclic graph with a node set $V(T)$ and an edge set $E(T)$. $T$
is a \textit{rooted tree} if exactly one node of $T$ has been designated the root. A node $v \in V(T)$ is a
\textit{leaf} if it has no child; otherwise, $v$ is an \textit{internal node}. A node $v \in V(T)$ is called in \textit{level} $i$, denoted by $level(v) = i$, means the number of edges on the path between the root and $v$ is $i$. Let $L(T)$ denote a subset of node set $V(T)$,
where each member is a leaf in $T$ and $n = |L(T)|$. Let \textit{heigth}$(T)$ denote the \textit{height} of tree $T$, which is max$\{level(v)|v \in L(T)\}$.
$T$ is a \textit{full binary tree} if each node of $T$
either has two children or it is a leaf. $T$ is a \textit{complete binary tree} if each internal node of $T$ has two children. Let $h= $ min$\{$height$(T_1), $ height$(T_2)\}$, and say height$(T_1) =h$ without loss of generality.

For a \textit{mixture tree} $T$, each leaf is associated with a species, and every internal node $v$ is associated
with a \textit{mutation time} $m_T(v)$ that represents the time when a mutation event occurs on the species node. In
fact, the mutation time of an internal node in a mixture tree can be regarded as the distance between the
node and any leaf of its descendants. Any two mixture tress $T_1$ and $T_2$ are \textit{comparable} if
$L(T_1) = L(T_2)$. Throughout this paper, the tree refers to a rooted full binary tree and
each internal node of the tree is associated with its mutation time, if not mentioned particularly.

Given any two nodes $u, v \in V(T)$, the \textit{least common ancestor} (abbreviated LCA) of $u$ and $v$ is
an ancestor of both $u$ and $v$ with the smallest mutation time. Let $P_T(u, v)$ denote the mutation time $m_T(w)$ of the
LCA $w$ of two leaves $u$ and $v$ in $T$. The \textit{mixture distance metric}, a metric for the mixture tree, is
formally defined as follows.\medskip

\noindent\textbf{Mixture distance metric.} The \textit{mixture distance} between two comparable mixture trees
$T_1$ and $T_2$, denoted by $d_m(T_1, T_2)$, is defined by the sum of difference of the mutation
times with respect to the LCAs of any two leaves in $T_1$ and $T_2$. That is, $d_m(T_1, T_2)$ =
\begin{math} \sum_{u, v \in L(T_1)=L(T_2)}|P_{T_1}(u, v) - P_{T_2}(u, v)|. \end{math}\\ \medskip

The significance of the mixture distance metric is to measure the similarity between two mixture trees, considering the mutation times (molecular clock) and mutation sites simultaneously. The paper is sought to develop two algorithms for efficiently computing the mixture distance between two
comparable mixture trees. Before we go into the algorithms, three properties of the mixture distance matric are demonstrated. Felsenstein\cite{Felsenstein2004} derived three mathematical properties -- reflexivity, symmetry and triangle inequality -- required for a well-defined metric. We show that the mixture distance is
well-defined in Theorem~\ref{metric_proterty}.

    \begin{thm}\label{metric_proterty}
    The mixture distance $d_m$ satisfies:
    \begin{enumerate}
    \item Reflexivity -- for any two comparable mixture trees $T_1$ and $T_2$, $d_m(T_1, T_2) = 0$ if and only if $T_1$ and $T_2$ are
    identical.
    \item Symmetry -- for any two comparable mixture trees $T_1$ and $T_2$, $d_m(T_1, T_2) = d_m(T_2, T_1)$.
    \item Triangle inequality -- for any three comparable mixture trees $T_1$, $T_2$ and $T_3$, $d_m(T_1, T_2) + d_m(T_2,
    T_3) \geq d_m(T_1, T_3)$.
    \end{enumerate}
    \end{thm}

    \begin{pf}
    Proof of 1. Due to $T_1 = T_2$, for any two nodes $u, v \in L(T_1)= L(T_2)$, we have $P_{T_1}(u, v) =
    P_{T_2}(u, v)$. Therefore, $d_m(T_1, T_2) = 0$ can be concluded. On the other hand, if $d_m(T_1, T_2) = 0$ for any two comparable mixture trees $T_1$ and $T_2$. We have $P_{T_1}(u, v) - P_{T_2}(u, v)$ for any $u, v \in L(T_1)=L(T_2)$ by the definition. Then we can prove $T_1 = T_2$ by induction on the height of $T_1$ (or $T_2$). \vspace*{0.2cm}

    Proof of 2. For any two nodes $u, v \in L(T_1) = L(T_2)$, $P_{T_1}(u, v) - P_{T_2}(u, v) = - (P_{T_2}(u, v) -
    P_{T_1}(u, v))$. Thus, $d_m(T_1, T_2) =$ $\sum_{u, v \in L(T_1)=L(T_2)}|P_{T_1}(u, v) -$ $P_{T_2}(u, v)| = \sum_{u, v \in L(T_1)=L(T_2)}|P_{T_2}(u, v) - P_{T_1}(u, v)|
    = d_m(T_2, T_1)$.\vspace*{0.2cm}

    Proof of 3. The triangle inequality is always satisfied for any three nonnegative number
    $a, b, c \in \Re^+ \cup {0}$, that is, $|a - b| + |b - c| \geq |a - c|$. Therefore,
    $|P_{T_1}(u, v) - P_{T_2}(u, v)| + |P_{T_2}(u, v) - P_{T_3}(u, v)| \geq |P_{T_1}(u, v) - P_{T_3}(u, v)|$
    is hold. Further, we have \vspace*{0.4cm}

    \hspace*{1.6cm}\begin{math} \sum_{u, v \in L(T_1)}|P_{T_1}(u, v) - P_{T_2}(u, v)| \end{math}\vspace*{0.2cm}

    \hspace*{1.6cm}\begin{math} + \sum_{u, v \in L(T_2)}|P_{T_2}(u, v) - P_{T_3}(u, v)| \end{math}\vspace*{0.2cm}

    \hspace*{1.6cm}\begin{math} \geq \sum_{u, v \in L(T_1)}|P_{T_1}(u, v) - P_{T_3}(u, v)| \end{math}.\vspace*{0.2cm}

    Consequently, $d_m(T_1, T_2) + d_m(T_2, T_3) \geq d_m(T_1, T_3)$ can be concluded. \qed
    \end{pf}

\section{An $O(n h)$-Time Algorithm}

Let $T_1$ and $T_2$ denote two comparable mixture trees of $n$ leaves for each tree. Note that, the mixture distance of $T_1$ and $T_2$ can be solved in $O(n^2)$-time: Because
when given two comparable mixture trees $T_1$ and $T_2$ each with $n$ leaves, there are $O(n^2)$
pairs of leaves separately in $T_1$ and $T_2$. In fact, the LCA of any pair of leaves can be found by adopting
the $O(1)$-time algorithm with $O(n)$-time preprocessing
~\cite{BF2000}.

In the following, another $O(n^2)$-time algorithm, named Algorithm~\textsc{MixtureDistance}, is proposed to compute the mixture
distance between $T_1$ and $T_2$. Which will help us to realize the next $O(nh)$-time algorithm, the main result.

Algorithm~\textsc{MixtureDistance} proceeds the nodes of $T_1$ by breadth-first search. For each internal
node $v$ in $T_1$, we find out the leaves of $T_1$ such that $v$ is exactly the LCA of each pair of the
leaves, and then compute the LCA $u$ of the leaves in $T_2$ which are mapped into the found leaves of $T_1$.
Finally, the difference of the mutation times between $u$ and $v$ is calculated. For convenience, we define $(a, b) * (c, d) = ad + bc$ for
any two ordered pairs $(a, b)$ and $(c, d)$.

\begin{figure*}[!h]
    \noindent\hrulefill\vspace{-0.6cm}
          \fontsize{10}{12pt}\selectfont
    \begin{tabbing}
    \hspace*{1em} \= \hspace*{1em} \= \hspace*{1em} \= \hspace*{1em} \= \hspace*{1em} \= \hspace*{1em} \= \hspace*{1em}
    \=
    \hspace*{1em} \kill \\
    \textbf{Algorithm} {\textsc{MixtureDistance}$(T_1, T_2)$}\\
    \textbf{Input:} Two comparable mixture trees $T_1$ and $T_2$, with mutation times $m_{T_1}(v)$ ($m_{T_2}(u)$,\\
     respectively) for every internal node $v$ of $T_1$ ($u$ of $T_2$, respectively).\\
    \textbf{Output:} The mixture distance $\mathcal{D}$ between $T_1$ and $T_2$.\\

    \ 1\> $\mathcal{D} = 0$.\\
    \ 2\> Traverse $T_1$ by the breadth-first search from its root and keep a list $\mathcal{I}_1$ of \\
       \> the internal nodes in order.\\
    \ 3\> Traverse $T_2$ by the breadth-first search from its root and keep a list $\mathcal{I}_2$ of \\
       \> the internal nodes in \textit{reverse} order.\\
    \ 4\> \textbf{for} each node $v \in \mathcal{I}_1$ \textbf{do}\\
    \ 5\>\> In $T_1$, color red the leaves of the left subtree rooted by $v$ and green the \\
       \>\> leaves of the right subtree rooted by $v$.\\
    \ 6\>\>\> \textbf{for} each node $u \in \mathcal{I}_2$ \textbf{do}\\
    \>\>\>\> // Initialize the coloring information of $u$'s children\\
    \ 7\>\>\>\> \textbf{for} each child $w$ of $u$ in $T_2$ \textbf{do}\\
    \ 8\>\>\>\>\> \textbf{if} $w$ is a leaf \textbf{then}\\
    \ 9\>\>\>\>\>\> \textbf{if} $w$ is colored by red in $T_1$ \textbf{then}\\
     10\>\>\>\>\>\>\> $color(w) = (1, 0)$.\\
     11\>\>\>\>\>\> \textbf{else if} $w$ is colored by green in $T_1$ \textbf{then}\\
     12\>\>\>\>\>\>\> $color(w) = (0, 1)$.\\
     13\>\>\>\>\>\> \textbf{else}\\
     14\>\>\>\>\>\>\> $color(w) = (0, 0)$.\\
     15\>\>\>\> Let $u_L$ and $u_R$ be the left and right children of $u$ in $T_2$, respectively.\\
    \>\>\>\> // Calculate the difference of the mutation times of $u$ and $v$ and \\
    \>\>\>\>\> sum them up for computing mixture distance\\
     16\>\>\>\> $number(u) = color(u_L) * color(u_R)$.\\
     17\>\>\>\> $\mathcal{D} = \mathcal{D} + |m_{T_1}(v) - m_{T_2}(u)| \times number(u)$.\\
    \>\>\>\> // Calculate the coloring information of $u$\\
     18\>\>\>\> $color(u) = color(u_L) + color(u_R)$.\\

    \end{tabbing}
    \vspace{-0.85 cm}\noindent\hrulefill
    \label{mixturetree}
    \end{figure*}

The algorithm adopts a 2-coloring method~\cite{BFP2003} on the leaves in $T_1$ and $T_2$ for easy
implementation. For each iteration associated with an internal node $v$ of $T_1$ in Line~4, the leaves
of the left and right subtrees rooted by $v$ are colored by red and green, respectively. The mapped leaves
in $T_2$ have the same coloring as one in $T_1$. The mixture distance between each internal node $u$ in $T_2$
and $v$ is calculated according the coloring scheme in $T_2$ (in Lines~16--17), and the coloring information
of $u$ would be derived for the computation of its parent node (in Line~18).

The coloring information of $u$, denoted by $color(u)$, indicates the coloring information of the subtree in
$T_2$ rooted by $u$. $color(u)$ includes two numbers of $u$'s descendant leaves colored by red
($color(u)[0]$) and green ($color(u)[1]$), respectively. $color(u)$ is derived by the coloring information of
its two children. That is, $color(u)[0] = color(u_L)[0] + color(u_R)[0]$ and $color(u)[1] = color(u_L)[1] +
color(u_R)[1]$, where $u_L$ and $u_R$ separately denote the left and right children of $u$ in $T_2$.

In Line~16, $number(u)$ is achieved by the special product of the color vectors of $u$'s two children.
$number(u) = color(u_L)[0] \times color(u_R)[1] + color(u_L)[1] \times color(u_R)[0]$. We multiply the
difference of their mutation times by $number(u)$ in Line~17, for computing the mixture distance between each
internal node $u$ in $T_2$ and $v$. At the end of Algorithm~\textsc{MixtureDistance}, $\mathcal{D}$ indicates
the mixture distance of $T_1$ and $T_2$.

After introducing Algorithm~\textsc{MixtureDistance}, we can give a $O(nh)$ computation time algorithm for computing the mixture distance between two mixture trees in the following part. In Algorithm~\textsc{MixtureDistance}, when the leaves of the subtree rooted by an internal node $v$ in $T_1$
are colored, other leaves in $T_1$ have no color, and so do the mapped leaves in $T_2$. That is, $color(w) =
(0, 0)$ for $w \in L(T_2)$. However, Algorithm~\textsc{MixtureDistance} still processes the ancestors of such leaves in $T_2$. In the following, we propose an algorithm for disregarding the nodes without
meaningful coloring information, and reduce the time complexity from $O(n^2)$ to $O(n h)$.

    The algorithm contains three main stages shown as follows:

    \begin{enumerate}\itemsep=2pt
    \item Rank the leaves in $T_1$ and $T_2$.
    \item Construct a minimal subtree $T_2'$ of $T_2$ involved in colored leaves with respect to node $v$,
    for each internal node $v$ in $T_1$.
    \item Compute the mixture distance between $v$ and each internal node in $T_2'$.
    \end{enumerate}

    \begin{figure}[h]\centering
    \includegraphics[scale=0.35]{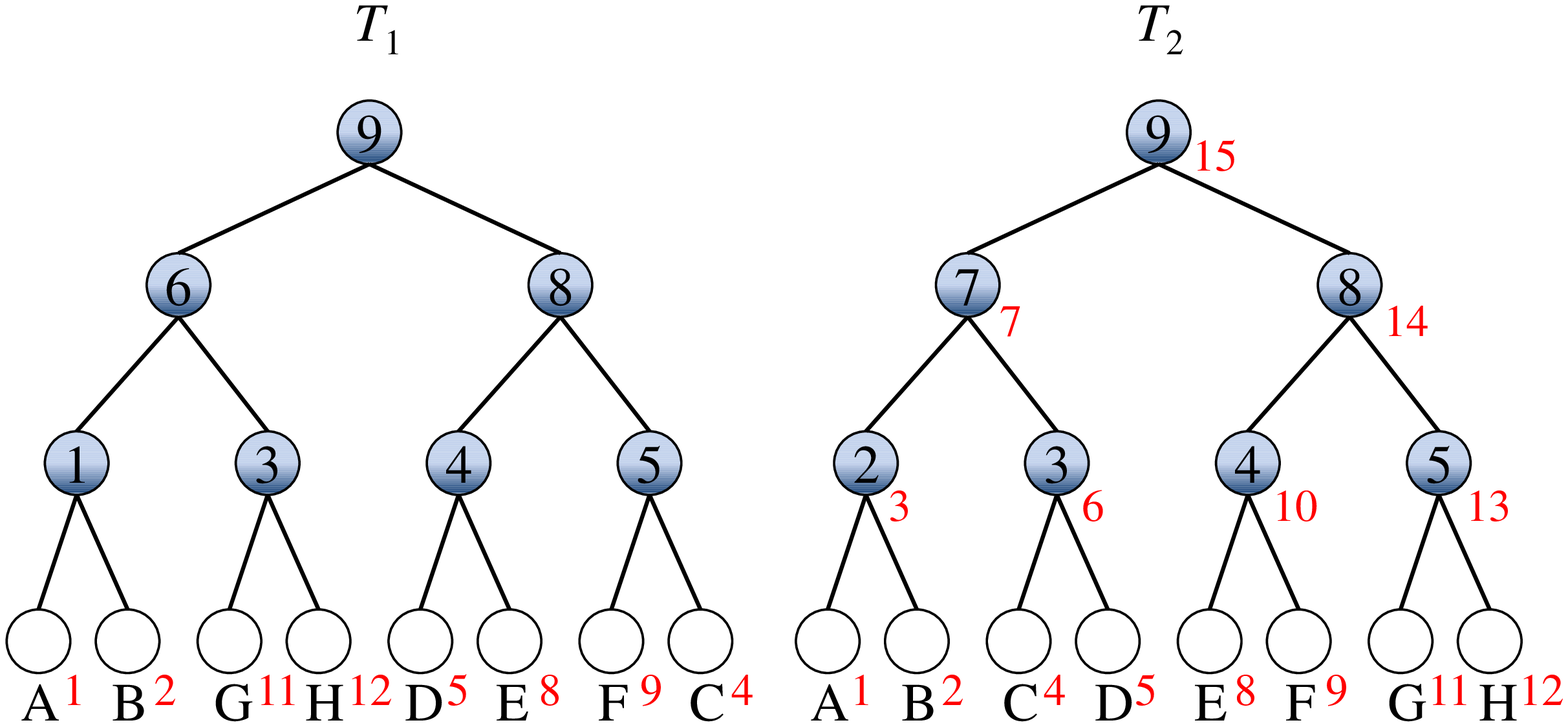}
    \caption{An example of ranking leaves of $T_1$ and $T_2$.}
        \label{fig_LeafLabeling}
    \end{figure}

In Stage~1, the nodes of $T_2$ is ranked in postorder, and the leaves of $T_1$ are assigned by the same rank
of the mapped leaves in $T_2$. In Fig.~\ref{fig_LeafLabeling}, red numbers nearby leaves in two given
comparable mixture trees $T_1$ and $T_2$ indicate the ranking achieved by Stage~1 of the algorithm.

The algorithm proceeds to Stage~2 for each internal node $v$ of $T_1$ in the reverse order of breadth-first
search. When $v$ in $T_1$ is processed, Stage~2 is sought to construct a minimal subtree $T_2'$ of $T_2$
involved in colored leaves with respect to node $v$. For node $v$, a nondecreasing list of the leaves of
the subtree rooted by $v$, denoted by $leaf(v)$, is obtained from the leaf lists of its two children, where
the leaves in the list are sorted by their ranks. Suppose that there are $k$ ordered nodes in $leaf(v)$,
that is, $leaf(v) = \{w_1, w_2, \ldots, w_k\}$. With the list $leaf(v)$, the subtree $T_2'$ can be
constructed as follows.

Let $lca(w_i, w_j)$ denote the LCA of leaves $w_i$ and $w_j$ in $T_2$, for any $i, j \in \{1, 2, \ldots, k\}$.
The subtree $T_2' = (V', E')$ is initialized by $V' = \{w_1, w_2, lca(w_1, w_2)\}$ and
$E' = \{\overline{lca(w_1, w_2)w_1}, \overline{lca(w_1, w_2)w_2}\}$. For node $w_i$,
$i \in \{1, 2, \ldots, k-2\}$,\vspace*{0.2cm}\\
\hspace*{1.6cm}\vspace*{0.2cm}$V' = V' \cup \{lca(w_{i+1}, w_{i+2}), w_{i+2}\}$ and\\
\hspace*{1.6cm}\vspace*{0.2cm}$E' = E' \cup \{\overline{lca(w_{i}, w_{i+1})lca(w_{i+1}, w_{i+2}}),
\overline{lca(w_{i+1}, w_{i+2})w_{i+2}}\}$.\vspace*{0.2cm}\\
Moreover, if the mutation time (the number written in the node circle) of $lca(w_{i}, w_{i+1})$ is larger
than the time of $lca(w_{i+1}, w_{i+2})$, the edge $\overline{lca(w_{i}, w_{i+1})w_{i+1}}$ is removed from
$E'$ and the edges $\overline{lca(w_{i+1}, w_{i+2})w_{i+1}}$ is inserted into $E'$.

    \begin{figure}[h]\centering
    \includegraphics[scale=0.35]{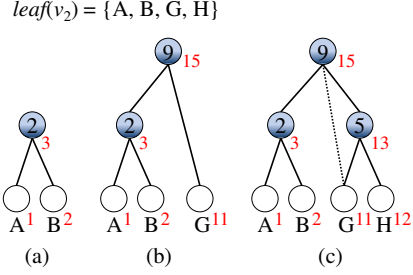}
    \caption{An example of constructing the subtree $T_2'$ with respect to $leaf(v_2)$ in Fig.~\ref{fig_LeafLabeling}.
        (a) The initialization of $T_2'$. (b) The intermediate of $T_2'$ as node A is processed. (c) The
        complete subtree $T_2'$ as node B is processed. Because the mutation time of $lca$(B, G) is
        larger than the time of $lca$(G, H), the dotted line incident to G is removed and the other
        incident edge of G is inserted.}
        \label{fig_SubtreeConstruction}
    \end{figure}

    \noindent\textit{Example 1.} An example of constructing the subtree $T_2'$ with respect to $leaf(v_2) = \{A, B, G, H\}$ is illustrated
    in Fig.~\ref{fig_SubtreeConstruction}. Initially, the node set $V'$ is $\{$A, B, $lca$(A, B)$\}$ and the
    edge set $E'$ includes the incident edges of the three nodes in $T_2$. As node A is processed, two
    nodes $lca$(B, G) and $G$ are inserted into $V'$, and two edges
    $\overline{lca(\mathrm{A}, \mathrm{B})lca(\mathrm{B}, \mathrm{G})}$ and
    $\overline{lca(\mathrm{B}, \mathrm{G})\mathrm{G}}$ are inserted into $E'$. Later, when node B is
    processed, two nodes $lca$(G, H) and H are inserted into $V'$ and two edges
    $\overline{lca(\mathrm{B}, \mathrm{G})lca(\mathrm{G}, \mathrm{H})}$ and
    $\overline{lca(\mathrm{G}, \mathrm{H})\mathrm{H}}$ are inserted into $E'$. Meanwhile, the edge
    $\overline{lca(\mathrm{B}, \mathrm{G})\mathrm{G}}$ is removed from $E'$ and the edge
    $\overline{lca(\mathrm{G}, \mathrm{H})\mathrm{G}}$ is inserted into $E'$, because the
    mutation time of $lca$(B, G) is larger than the time of the $lca$(G, H).\qed

After the subtree $T_2'$ with respect to currently processed node $v$ is constructed, Stage~3 of the
algorithm performs Lines~5--18 of Algorithm~\textsc{MixtureDistance} to computes the ``partial" mixture distance
between $T_2'$ and the subtree rooted by $v$ (only compute the distance of some nodes pairs which LCA is equal to $v$). At the end of the algorithm, $\mathcal{D}$ indicates the
mixture distance between $T_1$ and $T_2$.

    \begin{thm}\label{complexity2}
    The improved algorithm takes $O(n \log n)$ computation time and $O(n)$ preprocessing time, where $n$
    denotes the number of leaves of the mixture trees.
    \end{thm}

    \begin{pf}
    The algorithm contains three main stages. The first stage ranks the leaves in $T_1$ and $T_2$, which
    takes $O(n)$ time.

    In the second stage, a minimal subtree $T_2'$ of $T_2$ involved in colored leaves with respect to each
    node $v$ in $T_1$ is constructed. For each node $v$, a leaf list $leaf(v)$ is obtained from the leaf
    lists of its two children, which is achieved in $O(t)$ time by using the two-way merge
    algorithm~\cite{LCTT2005} performed in the leaf list of $v$'s children, where $t$ is the size of $leaf(v)$. The $O(1)$-time algorithm
    with $O(n)$-time processing~\cite{BF2000} is employed to compute the LCA of any pair of nodes in $T_2$.
    The last stage computes the mixture distance between $v$ and each internal node in $T_2'$ by performing
    Lines~5--18 of Algorithm~\textsc{MixtureDistance}, which takes $O(t)$ time. Although Stages~2 and~3 take
    $O(n)$ iterations in total. But each iteration deal with different  $t$ nodes. Note that for all internal nodes which in the same level of  $T_1$, the sum of $t$ (for each node) is $n$. Therefore, Stages~2 and~3 totally take $O(n h)$ time, where $h$ is the height of $T_1$. Hence, the algorithm requires $O(n h)$ computation time with
    $O(n)$ preprocessing time.\qed
    \end{pf}

\section{Conclusion}

In this paper, we provide a novel metric, named mixture distance metric, to measure the similarity between
two mixture trees.  It uniquely considers the estimated evolutionary time in the trees.
Two algorithms are developed to compute the mixture distance between mixture trees.  One requires $O(n^2)$ computation time and the other requires $O(nh)$ computation time with $O(n)$ preprocessing time, respectively. Note that when $T_1$ is a complete binary tree, $h$ will be $O(\log n)$ and the time complexity of our algorithm will be $(n \log n)$. In addition, we compare our approaches with the methods performed on
nodal distance metric~\cite{BS2003} and the results are shown in Table 
1. In shows our proposed approaches perform better than the methods performed on the nodal distance.

\begin{table}[h]
\label{table_1}
\caption{Metrics Comparison for Binary Trees}
\begin{center}
\begin{tabular}[h]{|c|p{2.8cm}|p{2.5cm}|p{2.5cm}|}
    \hline
    & & \multicolumn{2}{|c|}{Time complexity}\\
    \cline{3-4}
    \multicolumn{1}{|c}{Metric}&\multicolumn{1}{|c|}{Considerence}& Full binary trees & Complete binary trees\\
    \hline \hline
    Nodal distance& Structure & \multicolumn{1}{c|}{$O(n^3)$} &\multicolumn{1}{c|}{$O(n^2\log n)$} \\
    \hline
    Mixture distance& Structure and mutation time & \multicolumn{1}{c|}{\raisebox{-1.5ex}[0pt]{$O(nh)$}} &\multicolumn{1}{c|}{\raisebox{-1.5ex}[0pt]{$O(n\log n)$}}\\
    \hline

  \end{tabular}
\end{center}
\end{table}



\bibliographystyle{elsarticle-num}


\end{document}